\documentclass[aps,prc,twocolumn,superscriptaddress,nofootinbib,longbibliography,floatfix,10pt]{revtex4-1}

\usepackage{graphicx}
\usepackage{dcolumn}
\usepackage{bm}
\usepackage{morefloats}
\usepackage{multirow}
\usepackage{amssymb}
\usepackage{amsmath}
\usepackage{xcolor} 
\usepackage{longtable}
\usepackage{fix-cm}
\usepackage{mathptmx} 
\usepackage[T1]{fontenc}
\usepackage[colorlinks,allcolors=blue]{hyperref}
\makeatletter
\def\NAT@def@citea{\def\@citea{\NAT@separator}}
\makeatother

\begin{document}

\title{Heavy isotope production in ${}^{136}\text{Xe}+{}^{208}\text{Pb}$ collisions at $E_\text{c.m.}=514$~MeV}

\author{S. Ayik}\email{ayik@tntech.edu}
\affiliation{Physics Department, Tennessee Technological University, Cookeville, TN 38505, USA}
\author{O. Yilmaz}
\affiliation{Physics Department, Middle East Technical University, 06800 Ankara, Turkey}
\author{B. Yilmaz}
\affiliation{Physics Department, Faculty of Sciences, Ankara University, 06100 Ankara, Turkey}
\author{A. S. Umar}
\affiliation{Department of Physics and Astronomy, Vanderbilt University, Nashville, TN 37235, USA}

\date{\today}

\begin{abstract}
Employing the quantal diffusion mechanism for multi-nucleon transfer, the double differential cross-sections are calculated for production of primary projectile-like and target-like fragments in collisions of ${}^{136}\text{Xe}+{}^{208}\text{Pb}$ system at $E_\text{c.m.} =514$~MeV. Including de-excitation due to neutron emission, the cross-section for production of ${}^{210}\text{Po}$, ${}^{222}\text{Rn}$ and ${}^{224}\text{Ra}$ isotopes are estimated and compared with data.
\end{abstract}


\maketitle

\section{Introduction}
\label{sec1}
In recent years there has been a renewed interest for studying multi-nucleon transfer mechanism in dissipative heavy-ion collisions of massive nuclei at low energies. One of the reasons for these investigations is the possibility of producing heavy neutron rich nuclei by means of multi-nucleon transfer mechanism~\cite{dasso1994,corradi2009,sekizawa2019,kozulin2012,watanabe2015,desai2019}. The theoretical underpinnings of the multi-nucleon transfer processes are largely based on stochastic mechanisms of uncorrelated nucleon transfer between projectile-like and target-like nuclei during the collision.  These theoretical models can be broadly classified into the macroscopic and the microscopic approaches. In the macroscopic approach, macroscopic variables such as the mass and charge asymmetry, the relative distance, and the relative momentum are chosen from physical considerations and the evolution of these macroscopic variables are determined by means of phenomenological Langevin type semi-classical transport equations~\cite{zagrebaev2008c,zagrebaev2011,zagrebaev2012,karpov2017,saiko2019}. The approach involves a number of adjustable parameters which are determined under certain approximations. The time-dependent Hartree-Fock (TDHF) theory provides a microscopic basis for the description of the collision heavy-ion dynamics at low energies~\cite{negele1982,simenel2012,simenel2018}. However, TDHF theory has a fundamental limitation: it provides a good description of the most probable path of the collision dynamics including the one-body dissipation, but severely underestimates the fluctuations around the most probably path ~\cite{simenel2012,simenel2018,wakhle2014,oberacker2014,umar2015a,umar2016}. It is possible to calculate the dispersion of the mass and charge asymmetry distributions with the particle number projection method of the many-body TDHF wave function~\cite{sekizawa2016,sekizawa2017a}. Such an approach may give a reasonable description for a few nucleon transfers in quasi-elastic regime, but not in deep-inelastic and quasi-fission regimes. It is possible to calculate the dispersions of the one-body observables employing the time-dependent random-phase method developed by Balian and Veneroni ~\cite{balian1984,balian1985}. This method works well for symmetric collisions but is not applicable to asymmetric collisions~\cite{broomfield2008,broomfield2009,simenel2011,williams2018}. 

Recently, an extension of the TDHF approach beyond the mean-field description has been proposed, which is referred to as the stochastic mean-field (SMF) approach~\cite{ayik2008,lacroix2014}. In the SMF approach, instead of a single event an ensemble of mean-field events are generated by means of incorporating the fluctuations in the initial state. Alteranately, when di-nuclear structure is maintained during the collision, rather than generating an ensemble of events, it is possible to determine the dynamics of the collision in terms of  a few macroscopic variables such as mass and charge asymmetry, and the orbital angular momentum. The choice of the macroscopic variables is similar to the phenomenological approaches, but in the SMF approach it is possible to specify these macroscopic variables in terms of the TDHF single-particle states. The SMF approach gives rise to Langevin type stochastic transport equations for the evolution of these macroscopic variables. Furthermore the transport coefficients are determined by the occupied single-particle wave functions of the TDHF equations. The SMF approach includes the quantal effects due to shell structure and does not involve any adjustable variables other than the parameters of the Skyrme interaction. For the detailed description of the SMF approach and its reductions to macroscopic sub-space, we refer to~\cite{ayik2017,ayik2018,yilmaz2018,ayik2019}.

In a recent work, the quantal diffusion description based on the SMF approach was used to investigate nucleon transfer mechanism for the 
${}^{136}\text{Xe}+{}^{208}\text{Pb}$ system at $E_\text{c.m.}=526$~MeV~\cite{ayik2019}. In the present work, we investigate isotope production mechanism in the same system at slightly lower energy of $E_\text{c.m.} =514$~MeV. In Sec.~\ref{sec2}, we present calculations of the cross-section as a function of mass number and also the cross-section as a function of proton and neutron numbers in the vicinity of target-like primary fragments. In Sec.~\ref{sec3}, we present the estimated production cross-section of heavy-neutron rich target-like isotopes, 
${}^{210}\text{Po}$, ${}^{222}\text{Rn}$ and ${}^{224}\text{Ra}$ , after de-excitations by neutron emission. The conclusions are given in Sec.~\ref{sec4}.

\section{Neutron and proton distributions of the primary fragments in the collisions of ${}^{136}\text{Xe}+{}^{208}\text{Pb}$}
\label{sec2}
Recently, employing quantal diffusion description based on the SMF approach, we calculated the production cross-section $\sigma(A)$ of the primary fragments in the collisions of ${}^{136}\text{Xe}+{}^{208}\text{Pb}$ system at $E_\text{c.m.} =526$~MeV. As discussed in Ref.~\cite{ayik2018}, it is possible to extend this description to calculate cross section, $\sigma (N,Z)$, of production of the primary fragments as a function of neutron, $N$, and proton, $Z$, numbers
\begin{equation} \label{eq1}
\sigma (N,Z)=\frac{\pi \hbar ^{2} }{2\mu E_\text{c.m.} } \sum _{l_{\min } }^{l_{\max } }(2l+1)\left[P_{1,l} (N,Z)+P_{2,l} (N,Z)\right]\;,  
\end{equation} 
where the range of orbital angular momenta $\left(l_{\min },l_{\max }\right)$ is specified according to the angular position of detectors in the laboratory frame and the TKE cut-off. Here, $P_{1,l}(N,Z)$ and $P_{2,l}(N,Z)$ represent neutron and proton distribution functions for the projectile-like and target-like fragments, respectively and each distribution function is normalized to $0.5$, consequently the total probability is normalized to $1.0$. The distribution functions have the following correlated Gaussian form,
\begin{equation} \label{eq2} 
P_{l} (N,Z)=\frac{0.5}{2\pi \sigma _{NN} (l)\sigma _{ZZ} (l)\sqrt{1-\rho _{l}^{2} } } \exp \left[-C_{l} \left(N,Z\right)\right]\;.
\end{equation} 
Here, $\sigma _{NN}(l)$ and $\sigma _{ZZ}(l)$ are the dispersions of the neutron and proton distributions for each initial orbital angular momentum $l$, and the exponent $C_{l}\left(N,Z\right)$ for each angular momentum is given by
\begin{align} \label{eq3} 
&C_{l}\left(N,Z\right)=\frac{1}{2\left(1-\rho _{l}^{2} \right)}\nonumber\\
&\times\left[\left(\frac{Z-Z_{l} }{\sigma _{ZZ} (l)} \right)^{2} -2\rho _{l} \left(\frac{Z-Z_{l} }{\sigma _{ZZ} (l)} \right)\left(\frac{N-N_{l} }{\sigma _{NN} (l)} \right)+\left(\frac{N-N_{l} }{\sigma _{NN} (l)} \right)^{2} \right]\;,
\end{align} 
with the correlation factor $\rho_{l} =\sigma _{NZ}^{2} (l)/(\sigma_{ZZ} (l)\sigma_{NN} (l))$ determined by the ratio of the mixed variance to the product of neutron and proton dispersions. The quantities $N_{l}$ and $Z_{l}$ denote the mean values of the neutron and proton distribution of the projectile-like or the target-like fragments. The projectile-like and target-like primary fragment distribution functions for each angular momentum are specified by five quantities: neutron, proton, and mixed dispersions, and the mean value of neutrons and protons. The mean values of various quantities are determined directly from TDHF calculations. Table~\ref{tab1} provides, for the collision of ${}^{136}\text{Xe}+{}^{208}\text{Pb}$ system at $E_\text{c.m.}=514$~MeV, a list of the initial and final values of the mass numbers and proton numbers of projectile-like and target-like fragments, the initial and final angular momentum, the $TKE$ and $E^{*}$, as well as the scattering angles in the center of mass and laboratory frame over the angular momentum range $\left(l_{\min }=100,l_{\max }=280\right)$ at $\Delta l=20$ intervals.  This range is determined by the angular position of the detector in the laboratory frame and the cut-off of $E^*$ below $50$~MeV. 

Neutron, proton and mixed dispersions are determined by solving a set couple differential equations, given by Eqs.~(17-19) in Ref.~\cite{ayik2018}. The neutron, the proton, and the mixed diffusion coefficients as well as the derivatives of the neutron and the proton drift coefficients provide the input quantities for solving these equations. The expression for the diffusion coefficients are given by Eq.~(37) in Ref.~\cite{ayik2018}. The diffusion coefficients are determined in terms of the single-particle wave functions of the TDHF time evolution. The diffusion coefficients at $E_\text{c.m.} =514$~MeV, which a have similar behavior as those at $E_\text{c.m.} =526$~MeV, have slightly smaller magnitude and shorter contact times due to lower bombarding energy. We can specify the derivatives of the drift coefficients with the help of the mean drift-path in the $\left(N,Z\right)$ plane.  Symmetric systems and those systems with strong shell effects, on the average, do not exhibit nucleon drift between projectile-like fragments and target-like fragments. In these cases, it is possible to determine the derivative of the drift coefficients by calculating the TDHF evolution of a suitable neighboring system which is driven towards the equilibrium position of the system under investigation. The ${}^{136}\text{Xe}+{}^{208}\text{Pb}$ system is an example of such a system. Lead is a doubly closed shell nucleus with neutron and proton numbers are $N=126$, $Z=82$ and the neutron shell is closed in xenon with $N=82$.  As a result of strong shell effects the system is located at a local minimum position in the potential energy surface in the $\left(N,Z\right)$ plane. As seen from Table~\ref{tab1}, the system exhibits only a small mass drift on the order one nucleon. With such a small drift, it is not possible to determine the derivatives of the drift coefficients. As illustrated in Fig.~3 of Ref.~\cite{ayik2019}, we refer to the line in the 
$\left(N,Z\right)$ plane which joins the position of ${}^{136}\text{Xe}$ with the position of ${}^{208}\text{Pb}$ as the iso-scalar path, and the line perpendicular to the iso-scalar path is referred to as the iso-vector path.
\begin{table}[h]
\caption{Initial and final values of the mass and proton numbers of projectile-like and target-like fragments, the initial and the final angular momentum, the final $TKE$ and the excitation $E^*$, scattering angles in the center of mass and laboratory frames over the angular momentum range $l_{i} =100-280$ at $\Delta l=20$ intervals in the collisions of ${}^{136}\text{Xe}+{}^{208}\text{Pb}$ at $E_\text{c.m.} =514$~MeV.}
\label{tab1}
\begin{ruledtabular}
\begin{tabular}{c c c c c c c c c c c }
$\ell_i\,$($\hbar$) & A$_1^f$ & Z$_1^f$ & A$_2^f$ & Z$_2^f$ & $\ell_f\,$($\hbar$) & TKE  
& $\theta_{c.m.}$ & E$^*$ & $\theta_1^{lab}$ & $\theta_2^{lab}$ \\
&&&&&&(MeV)&&(MeV)&& \\
\hline
\rule{0pt}{3ex}
100	& 136 & 54.1 & 208 & 81.9 &	83.3 & 353 & 126 & 161 & 76.1 &	24.2\\
120	& 137 &	54.6 & 207 & 81.4 &	101 & 354 &	117 & 160 &	68.8 &	28.6\\
160	& 140 &	55.5 & 204 & 80.5 &	132	& 356 &	99.4 & 158 & 57.1 &	36.4\\
180	& 139 &	55.2 & 205 & 80.8 &	150 & 350 &	92.1 & 160 & 52.5 &	39.1\\
200	& 137 &	54.8 & 207 & 81.2 &	161 & 347 &	87.6 & 168 & 49.8 &	40.6\\
220	& 138 &	55.5 & 206 & 80.5 &	178	& 362 &	85.2 & 149 & 48.7 &	42.4\\
240	& 138 &	55.4 & 206 & 80.6 &	197 & 386 &	83.9 & 125 & 48.9 &	43.8\\
260	& 136 &	54.8 & 208 & 81.2 &	222 & 415 &	82.7 & 94.0 & 49.2 & 45.2\\
280	& 137 &	54.7 & 207 & 81.4 &	254 & 455 &	82.0 & 58.8 & 49.8 & 47.1\\
\end{tabular}
\end{ruledtabular}
\end{table}
In the previous study of the ${}^{136}\text{Xe}+{}^{208}\text{Pb}$ system at $E_\text{c.m.} =526$~MeV, we determined the curvatures of the local potential energy at the equilibrium point by following the mean trajectories in the collisions of neighboring systems ${}^{130}\text{Te}+{}^{214}\text{Po}$ and ${}^{138}\text{Ce}+{}^{206}\text{Pt}$. As discussed in that work, the derivative of the drift coefficients are closely related to the curvature of the potential energy surface at the equilibrium position of ${}^{136}\text{Xe}$ (or ${}^{208}\text{Pb}$) along the iso-scalar and iso-vector directions. Since the potential energy surface should not strongly depend on the bombarding energy, we use the same curvature parameters obtained at $E_\text{c.m.} =526$~MeV in the collision at lower bombarding energy of $E_\text{c.m.} =514$~MeV. Furthermore, as a consequence of strong shell effects, the potential energy at the minimum location of ${}^{136}\text{Xe}$ (or ${}^{208}\text{Pb}$)  in the iso-scalar direction has different curvatures toward symmetry and asymmetry. As a result, the magnitudes of dispersions have slightly larger values in symmetry direction $\sigma _{NN}^{>} (l),\sigma _{ZZ}^{>} (l),\sigma _{NZ}^{>} (l)$  than in asymmetry direction $\sigma _{NN}^{<} (l),\sigma _{ZZ}^{<} (l),\sigma _{NZ}^{<} (l)$, as well. Table~\ref{tab2} illustrates the final values of the neutron variances, proton variances, mixed variances and the variances of the total mass numbers over the range of initial angular momentum $\left(l_{\min } =100,l_{\max } =280\right)$ at the end of each $\Delta l=20$ interval. In Table~\ref{tab2}, dispersion of the mass distributions toward symmetry and asymmetry directions are determined using the expression $\sigma _{AA}^{2} (l)=\sigma _{NN}^{2} (l)+\sigma _{ZZ}^{2} (l)+2\sigma _{NZ}^{2} (l)$ for the range of initial angular momenta. 
\begin{table}[h]
\caption{Neutron, proton and mixed variances, mass dispersions toward asymmetry and symmetry directions and the dispersions of the middle Gauss functions over the angular momentum range $l_{i} =100-280$ at $\Delta l=20$ intervals in the collisions of ${}^{136}\text{Xe}+{}^{208}\text{Pb}$ at $E_\text{c.m.} =514$~MeV.}
\label{tab2}
\begin{ruledtabular}
\begin{tabular}{ c c c c c c c c c c}
$\ell_i\,$($\hbar$) & $\sigma_{NN}^{2<}$ & $\sigma_{ZZ}^{2<}$ & $\sigma_{NZ}^{2<}$ & $\sigma_{AA}^<$ & $\sigma_{NN}^{2>}$ & $\sigma_{ZZ}^{2>}$ & $\sigma_{NZ}^{2>}$ & $\sigma_{AA}^>$ & $\overline{\sigma}_{AA}$ \\
\hline
\rule{0pt}{3ex}
100 & 26.1 & 19.9 &	0.70 &	6.89 &	58.4 &	28.9 &	17.7 &	11.1 &	8.98\\
120	& 25.5 & 19.3 &	0.69 &	6.80 &	57.0 &	28.0 &	17.1 &	10.9 &	8.86\\
140	& 24.9 & 18.8 &	0.68 &	6.71 &	55.3 &	27.0 &	16.5 &	10.7 &	8.73\\
160 & 24.1 & 18.2 &	0.66 &	6.60 &	52.8 &	25.9 &	15.5 &	10.5 &	8.54\\
180 & 23.5 & 17.6 &	0.63 &	6.51 &	49.7 &	24.4 &	14.1 &	10.1 &	8.31\\
200 & 22.4 & 16.4 &	0.57 &	6.32 &	45.5 &	22.2 &	12.2 &	9.59 &	7.96\\
220 & 19.6 & 14.2 &	0.49 &	5.97 &	38.6 &	18.7 &	9.79 &	8.77 &	7.37\\
240 & 16.1 & 11.7 &	0.39 &	5.34 &	29.9 &	14.7 &	6.88 &	7.64 &	6.49\\
260 & 12.9 & 8.71 &	0.27 &	4.70 &	21.5 &	10.2 &	3.87 &	6.33 &	5.52\\
280 & 9.08 & 5.21 &	0.13 &	4.10 &	12.9 &	5.63 &	1.40 &	4.62 &	4.36
\end{tabular}
\end{ruledtabular}
\end{table}

\subsection{Mass distribution of primary fragments}
\begin{figure}[!hpt]
\includegraphics*[width=8.6cm]{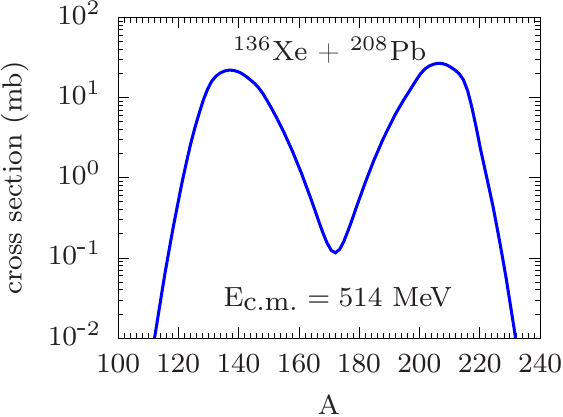}
\caption{(color online) Cross-sections $\sigma (A)$ for production of primary fragments as a function of mass number in the collisions of ${}^{136}\text{Xe}+{}^{208}\text{Pb}$ at $E_\text{c.m.}=514$~MeV.}
\label{fig1}
\end{figure}

We can calculate the probability distribution of the primary fragments as function of mass number by integrating Eq.~\eqref{eq2} over neutron and proton numbers while keeping the total mass number of the projectile-like and target-like fragments constant. The mass number distributions have Gaussian forms with different dispersions toward symmetry and asymmetry,
\begin{equation} \label{eq4} 
P_{l}^{<} (A-A_{l} )=\frac{0.5}{\sqrt{2\pi } } \frac{1}{\sigma _{AA}^{<} (l)} \exp \left[-\frac{1}{2} \left(\frac{A-A_{l} }{\sigma _{AA}^{<} (l)} \right)^{2} \right] 
\end{equation} 
and 
\begin{equation} \label{eq5} 
P_{l}^{<} (A-A_{l} )=\frac{0.5}{\sqrt{2\pi } } \frac{1}{\sigma _{AA}^{<} (l)} \exp \left[-\frac{1}{2} \left(\frac{A-A_{l} }{\sigma _{AA}^{<} (l)} \right)^{2} \right]\;,
\end{equation} 
where $A_{l}$ denotes the mean mass number of the projectile-like or the target-like fragments. We note that the distribution functions do not match at the mean value of the mass number $A=A{}_{l}$. The discontinuous behavior of distribution functions at the mean value arises from the discontinuity of the derivatives of the drift coefficients at the location of ${}^{136}\text{Xe}$ (or ${}^{208}\text{Pb}$).  As discussed in our recent article~\cite{ayik2019}, it is possible to provide an approximate description of the mass distribution function by smoothly joining the right and left Gauss functions by a middle Gauss function. For this purpose, we determine the left $A_{l}^{0} (L)$ and right $A_{l}^{0} (R)$ intersection points of the left and right Gauss functions from the condition $P_{l}^{<} (A_{l}^{0} -A_{l})=P_{l}^{<} (A_{l} -A_{l}^{0} )=\bar{P}_{l}$ and define the middle Gauss functions as,
\begin{equation} \label{eq6} 
\bar{P}_{l} (A-A_{l} )=\bar{P}_{l} \exp \left[-\frac{1}{2} \left(\frac{A-A_{l} }{\bar{\sigma }_{AA} (l)} \right)^{2} +\frac{1}{2}\;, \left(\frac{\Delta A_{l} }{\bar{\sigma }_{AA} (l)} \right)^{2} \right]\;,
\end{equation} 
where $\Delta A_{l} =A_{l} -A_{l}^{0} (L)=A_{l}^{0}(R)-A_{l}$. The dispersions $\bar{\sigma}_{AA} (l)$ of the middle Gauss functions are determined by requiring the entire distribution is normalized to $0.5$ for each orbital angular momentum. This requirement is given by Eq.~(34) in Ref.~\cite{ayik2019}. The last column in Table~\ref{tab1} shows the dispersions $\bar{\sigma }_{AA} (l)$ of the middle Gauss functions for the range of angular momenta. Also, we note that the dispersion of the middle Gauss functions is approximately equal to the average values of mass dispersions towards the asymmetry and asymmetry directions, $\bar{\sigma }_{AA} (l)\approx \left(\sigma _{AA}^{<} (l)+\sigma _{AA}^{>} (l)\right)/2$.  In a manner similar to Eq.~\eqref{eq1}, we calculate the cross-section for production of a fragment with mass number $A$ in the collision of ${}^{136}\text{Xe}+{}^{208}\text{Pb}$ at lower bombarding energy $E_\text{c.m.} =514$~MeV as,
\begin{equation} \label{eq7} 
\sigma (A)=\frac{\pi \hbar ^{2} }{2\mu E_\text{c.m.}} \sum _{l=100}^{l=280}(2l+1)\left[P_{1,l} (A)+P_{2,l} (A)\right]\;.
\end{equation} 
Here, $P_{1,l} (A)$ and $P_{2,l} (A)$ denote the probability distribution functions of the projectile-like and the target-like fragments for each initial orbital angular momentum $l$.  Probability distributions are evaluated at $\Delta l=20$ intervals and smoothly interpolated for integer values of the angular momenta. The distribution functions of the projectile-like fragments and the target-like fragments are determined according to,
\begin{equation} \label{eq8} 
P_{1,l} (A)=\left\{\begin{array}{ccc} {P_{l}^{<} (A-A_{1l} )} & {A\le A_{1l}^{0} (L)} & {} \\ {\bar{P}_{l} (A-A_{1l} )} & {A_{1l}^{0} (L)\le A\le A_{1l}^{0} (R)} & {} \\ {P_{l}^{>} (A-A_{1l} )} & {A\ge A_{1l}^{0} (R)} & {} \end{array}\right.  
\end{equation} 
and
\begin{equation} \label{eq9} 
P_{2,l} (A)=\left\{\begin{array}{ccc} {P_{l}^{<} (A-A_{2l} )} & {A\le A_{2l}^{0} (L)} & {} \\ {\bar{P}_{l} (A-A_{2l} )} & {A_{2l}^{0} (L)\le A\le A_{2l}^{0} (R)} & {} \\ {P_{l}^{>} (A-A_{2l} )} & {A\ge A_{2l}^{0} (R)} & {} \end{array}\right.\;.
\end{equation} 
In these expressions $A_{1l}$ and $A_{2l}$ denote the mean values of the mass numbers of the projectile-like and target-like fragments. Fig.~\ref{fig1} shows the cross-section in the collision of ${}^{136}\text{Xe}+{}^{208}\text{Pb}$ at $E_\text{c.m.} =514$~MeV for production of the primary fragments as function of the mass number $A$.  This figure is very similar to the cross sections given by Fig.~8 in Ref.~\cite{ayik2019} at slightly higher energy $E_\text{c.m.} =526$~MeV. There is no data provided at $E_\text{c.m.} =514$~MeV collision in 
Ref.~\cite{kozulin2012}.  We note that because of the larger dispersion towards symmetry, the cross-sections are not symmetric around the mean values, but the system has a tendency to diffuse toward the symmetry direction.

\subsection{Mass and charge distributions of primary fragments}
\begin{figure}[!hbt]
\includegraphics*[width=8.6cm]{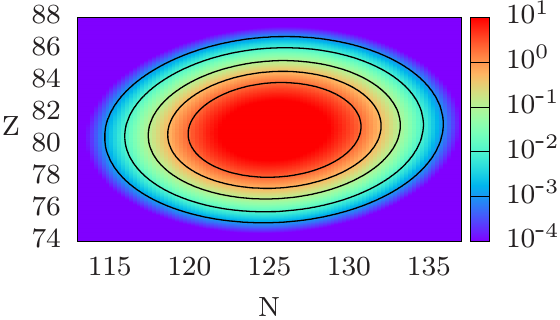}
\caption{(color online) Cross-section $\sigma (N,Z)$ for production of primary fragments with neutron $N$ and proton $Z$ numbers in $(N,Z)$-plane in the collisions of ${}^{136}\text{Xe}+{}^{208}\text{Pb}$ at $E_\text{c.m.} =514$~MeV. Elliptic lines indicate the set of isotopes with equal production cross-sections.}
\label{fig2}
\end{figure}

Using Eq.~\eqref{eq1}, we can calculate the production cross-sections in the collision of ${}^{136}\text{Xe}+{}^{208}\text{Pb}$ as functions of neutron and proton numbers of the primary fragments. Because of different neutron, proton and mixed dispersions toward symmetry and asymmetry directions of the iso-scalar path, the distributions functions have discontinuity at the mean positions of the projectile-like and target-like fragments. By extending the one dimensional description for the mass number $A$ distribution discussed in the previous section to two dimensions in the $\left(N,Z\right)$-plane, it is possible to determine a continuous  distribution function $P_{l} (N,Z)$ by smoothly joining the right $P_{l}^{>} (N,Z)$ and left $P_{l}^{<} (N,Z)$ distribution functions by a middle distribution $\bar{P}_{l} (N,Z)$ function  for each angular momentum. Here, rather than following this treatment, we employ a simpler description. We determine the smooth distribution function of the projectile-like and target-like fragments approximately by taking the average values of dispersions toward symmetry and asymmetry directions for each initial angular momentum,
\begin{equation} \label{eq10} 
P_{l} (N,Z)\approx \frac{0.5}{2\pi \bar{\sigma }_{NN} (l)\bar{\sigma }_{ZZ} (l)\sqrt{1-\bar{\rho }_{l}^{2} } } \exp \left[-\bar{C}_{l} \left(N,Z\right)\right]\,.
\end{equation} 
Here, the exponent $\bar{C}_{l} \left(N,Z\right)$ for each initial orbital angular momentum is given by
\begin{align} \label{eq11} 
&\bar{C}_{l}\left(N,Z\right)=\frac{1}{2\left(1-\bar{\rho }_{l}^{2} \right)}\nonumber\\
&\times \left[\left(\frac{Z-Z_{l} }{\bar{\sigma }_{ZZ} (l)} \right)^{2} -2\bar{\rho }_{l} \left(\frac{Z-Z_{l} }{\bar{\sigma }_{ZZ} (l)} \right)\left(\frac{N-N_{l} }{\bar{\sigma }_{NN} (l)} \right)+\left(\frac{N-N_{l} }{\bar{\sigma }_{NN} (l)} \right)^{2} \right] 
\end{align} 
with $\bar{\rho }_{l} =\bar{\sigma }_{NZ}^{2} (l)/(\bar{\sigma }_{ZZ} (l)\bar{\sigma }_{NN} (l))$. The quantities $\bar{\sigma }_{NN} (l)=\left(\sigma _{NN}^{>} (l)+\sigma _{NN}^{<} (l)\right)/2$, $\bar{\sigma }_{ZZ} (l)=\left(\sigma _{ZZ}^{>} (l)+\sigma _{ZZ}^{<} (l)\right)/2$ and $\bar{\sigma }_{NZ} (l)=\left(\sigma _{NZ}^{>} (l)+\sigma _{NZ}^{<} (l)\right)/2$ represent the average values of dispersions toward symmetry and asymmetry directions, and $N_{l}$ and $Z_{l}$ denote the mean neutron and proton numbers of the projectile-like or target-like fragments at each initial angular momentum. Employing the expression Eq.~\eqref{eq1}, we can calculate the cross-sections for producing primary projectile-like and target-like isotopes with mass number $A=N+Z$ with atomic number $Z$.  Figure~\ref{fig2} shows contour plots of production cross-sections for primary target-like nuclei in units of millibarn. Because of the approximate description of the smoothed probability distributions, these  cross-sections provide a good approximation for producing heavy isotopes in the neighborhood of the target nucleus. 
 
\section{Heavy isotope production in the collisions of ${}^{136}\text{Xe}+{}^{208}\text{Pb}$}
\label{sec3}
\begin{figure}[!hpt]
\includegraphics*[width=8.6cm]{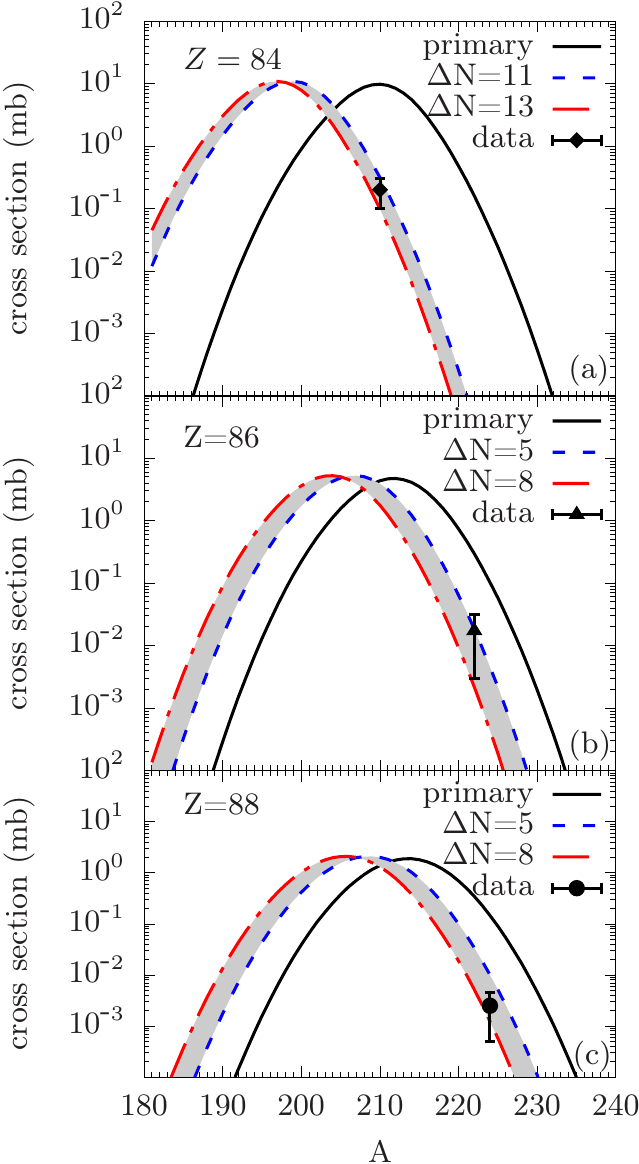}
\caption{(color online) Solid lines are the cross-sections in units of millibarn for production of primary isotopes of elements with atomic numbers $Z=84$, $Z=86$ and $Z=88$ plotted as a function of mass number. Shaded areas show the boundaries of the cross sections after de-excitation by emission of neutrons within the ranges $\Delta N=11-13$ for $Z=84$ and $\Delta N=5-8$ for $Z=86$, $Z=88$ isotopes.  Solid diamond, triangle and circle represent the data for ${}^{210}\text{Po}$,  ${}^{222}\text{Rn}$  and ${}^{224}\text{Ra}$. Taken from Ref.~\cite{kozulin2012}.}
\label{fig3}
\end{figure}

In the collisions of ${}^{136}\text{Xe}+{}^{208}\text{Pb}$ at $E_\text{c.m.} =514$~MeV the production cross-sections for heavy isotopes ${}^{210}\text{Po}$, ${}^{222}\text{Rn}$ and ${}^{224}\text{Ra}$ are reported in Ref.~\cite{kozulin2012}. In Fig.\ref{fig3} the measured cross-sections are indicated by solid diamond, triangle, and circle for ${}^{210}\text{Po}$, ${}^{222}\text{Rn}$, and ${}^{224}\text{Ra}$, respectively.  The primary fragments produced in the collision de-excite by particle emissions, mostly neutrons, and secondary fission of heavy fragments. For an accurate description of the de-excitation processes, we need to have information about the excitation energy and spin angular momentum of the primary fragments. The TDHF calculations presented in Table~\ref{tab1} give the mean total excitation energy and final total orbital angular momentum of the fragments. 
For the moment we do not have accurate information about the excitation energy, and division of the excitation energy and the spin angular momentum of the final fragments. Here, we provide an estimate of the isotope cross-sections on the heavy side of the target nucleus in the following manner: It is reasonable to determine the distribution functions of heavy primary isotopes to the right of the target nucleus in terms of the dispersions values toward the asymmetry direction as,
\begin{equation} \label{eq12} 
P_{2,l}^{<} (N,Z)\approx \frac{0.5\;\exp{\left[-C_{2,l}^{<} \left(N,Z\right)\right]}}{2\pi \sigma _{NN}^{<} (l)\sigma _{ZZ}^{<} (l)\sqrt{1-(\rho _{l}^{<} )^{2} } } \;.
\end{equation} 
Here the exponent $C_{l}^{<}\left(N,Z\right)$ for each initial orbital angular momentum is given by
\begin{align} \label{eq13} 
C_{2,l}^{<} \left(N,Z\right)=\frac{1}{2\left(1-(\rho _{l}^{<} )^{2} \right)}&\left[\left(\frac{Z-Z_{2l} }{\sigma _{ZZ}^{<} (l)} \right)^{2} +\left(\frac{N-N_{2l} }{\sigma _{NN}^{<} (l)} \right)^{2} \right.\nonumber\\
&\left.-2\rho _{l}^{<} \left(\frac{Z-Z_{2l} }{\sigma _{ZZ}^{<} (l)} \right)\left(\frac{N-N_{2l} }{\sigma _{NN}^{<} (l)} \right)\right]
\end{align} 
with  $\rho _{l}^{<} =(\sigma _{NZ}^{<} (l))^2/\sigma _{ZZ}^{<} (l)\sigma _{NN}^{<} (l)$ and $N_{2l} $, $Z_{2l} $ indicate the mean values of neutron and proton numbers of the target-like fragments. Using Eq~\eqref{eq1} with the probability distributions~\eqref{eq12}, we calculate the cross-sections of producing primary isotopes on the heavy side of the target nucleus as,
\begin{equation} \label{eq14}
\sigma _{2}^{<} (N,Z)=\frac{\pi \hbar ^{2} }{2\mu E_\text{c.m.} } \sum _{l=100}^{l=280}(2l+1)P_{2,l}^{<} (N,Z)\;.
\end{equation} 
Solid lines in Fig.~\ref{fig3} show the cross-section for producing primary isotopes $Z=84$ (a), $Z=86$ (b) and $Z=88$ (c) as a function of the mass number. We assume that the heavy neutron rich primary fragments de-excite mainly by neutron emission. We calculate the secondary 
cross-sections after neutron emission using the same expression by replacing $\sigma _{2}^{<} (N,Z)\to \sigma _{2}^{<} (N+\Delta N,Z)$, 
where $\Delta N$ denotes the number of neutrons emitted from the primary fragment with the total number of neutrons $\Delta N+N$. Since we don't have accurate information about the excitation energies of the primary fragments, we do not have accurate information about the numbers of emitted neutrons. For an estimation of the cross-sections, we take a range of emitted neutrons $\Delta N=11-13$ for the polonium isotopes, $\Delta N=5-8$ for the radon and the radium isotopes. These ranges for numbers of emitted neutrons approximately correspond to excitation energies of the primary fragments $E^*\approx 100$~MeV and $E^*\approx 60$~MeV with neutron numbers $N+\Delta N$, respectively. From Table~\ref{tab1}, it is also possible to estimate excitation energies of these heavy primary isotopes which are produced mainly in the angular momentum range of $l_{i} =100-220$. We can approximately determine the excitation energies of the neutron rich primary, polonium, radon and radium isotopes by taking into account the ground state $Q_{gg}$-value corrections relative to the entrance channel ${}^{136}\text{Xe}+{}^{208}\text{Pb}$ and by sharing the excitation energies in proportion to the masses of the binary primary fragments. These estimations are consistent with the excitation energies approximated from the range of emitted neutrons from each primary isotope. The estimated cross-sections of the secondary isotopes are indicated by the gray area in Fig.~\ref{fig3}.  Data for the ${}^{210}\text{Po}$, ${}^{222}\text{Rn}$ and${}^{224}\text{Ra}$ nuclei, with experimental error bars, are located in the gray area for each isotope. Our estimates of the measure cross-sections are within the range of data.

\section{Conclusions}
\label{sec4}
Employing the quantal diffusion description based on the SMF approach, we have investigated multi-nucleon transfer mechanism in the collisions of ${}^{136}\text{Xe}+{}^{208}\text{Pb}$ at $E_\text{c.m.} =514$~MeV. Extending our previous description carried out for the same system at $E_\text{c.m.} =514$~MeV, in additions to the cross-section as a function of the mass numbers, we calculate cross-sections for production of the primary isotopes as a functions neutron and proton numbers. Incorporating the de-excitation mechanism by neutron emission in an approximate manner, we provide estimates of the cross-sections for production of the heavy isotopes the ${}^{210}\text{Po}$, ${}^{222}\text{Rn}$ and ${}^{224}\text{Ra}$ nuclei and compare with data. The estimated cross-sections are within the range of the measured data. We plan to improve the cross-section calculations further by determining the excitation energies of the heavy primary fragments more accurately including the fluctuations of the excitation energy.   
 
\begin{acknowledgments}
S.A. gratefully acknowledges the IPN-Orsay and the Middle East Technical University for warm hospitality extended to him during his visits. S.A. also gratefully acknowledges useful discussions with D. Lacroix, and very much thankful to F. Ayik for continuous support and encouragement. This work is supported in part by US DOE Grants No. DE-SC0015513,  in part by US DOE Grant No. DE-SC0013847, and  in part TUBITAK Grant No. 117F109.
\end{acknowledgments}
 
\bibliography{VU_bibtex_master}
\end{document}